\documentclass[twocolumn,epjc3,a4paper]{svjour3}
\usepackage[numbers,sort&compress]{natbib}

\usepackage[utf8]{inputenc}
\usepackage[english]{babel}
\usepackage{amssymb,amsmath,amstext,amsbsy,amsopn}
\usepackage{bbm}
\usepackage{graphicx}
\usepackage{tabularx}
\usepackage{lmodern}

\usepackage{microtype}

\journalname{Eur. Phys. J. A}

\usepackage[
colorlinks=true,
linkcolor=blue,
citecolor=blue,
urlcolor=blue
]{hyperref}
\usepackage{orcidlink} 

\newcommand{\Qsq}{\ensuremath{Q^{2}}}
\newcommand{\GeV}{\ensuremath{\,\mathrm{GeV}}}
\newcommand{\Fone}{\ensuremath{F_{1}}}
\newcommand{\Ftwo}{\ensuremath{F_{2}}}
\newcommand{\GE}{\ensuremath{G_{E}}}
\newcommand{\GM}{\ensuremath{G_{M}}}

\newcommand{\twopi}{\ensuremath{2\pi}}
\newcommand{\threepi}{\ensuremath{3\pi}}
\newcommand{\qqbar}{\ensuremath{q\bar q}}

\begin{document}
\allowdisplaybreaks

\title{Nucleon electromagnetic form-factors in a minimal Gari--Kr{\"u}mpelmann model including
	two-meson continua}

\author{\mbox{Wulf  Kr\"umpelmann}\thanksref{add1,e1} \orcidlink{0009-0006-8745-4305}\and \nolinebreak[4]
\mbox{Yong-Hui Lin}\thanksref{add2,e2}\orcidlink{0000-0001-8800-9437} \and \nolinebreak[4]
\mbox{Ulf-G.~Mei{\ss}ner}\thanksref{add4,add5,e3}\orcidlink{0000-0003-1254-442X} }

\thankstext{e1}{e-mail:~wulf.kruempelmann@gmail.com}
\thankstext{e2}{e-mail:~yonghuil@buaa.edu.cn}
\thankstext{e3}{e-mail:~meissner@hiskp.uni-bonn.de}

\institute
{Helmholtz-Institut f\"ur Strahlen- und Kernphysik, Universit\"at Bonn,
	D-53115 Bonn, Germany \label{add1}
	\and
	Peng Huanwu Collaborative Center for Research and Education, International Institute for Interdisciplinary
	and Frontiers, Beihang University, Beijing 100191, China\label{add2}
	\and
	Helmholtz-Institut f\"ur Strahlen- und Kernphysik, Bethe Center for Theoretical Physics and Cluster of Excellence - Color meets Flavor, Universit\"at Bonn,
	D-53115 Bonn, Germany \label{add4}
	\and
	Institute for Advanced Simulation (IAS-4), Forschungszentrum J\"ulich, D-52425 J\"ulich, Germany \label{add5}
}

\date{}
\maketitle

\begin{abstract}
The Gari--Kr\"umpelmann (GK) model provided one of the first semiphenomenological descriptions of the
electromagnetic nucleon form-factors over a wide range of momentum transfer. It combined vector-meson
dominance at low and intermediate momentum transfer with perturbative-QCD constraints at large momentum
transfer. The present work extends this framework by including the dispersive two-pion continuum explicitly
in the isovector channel. The two-pion contribution is taken from modern dispersive analyses and embedded
into the intrinsic high-\Qsq{} structure of the GK ansatz.
The physical $\rho$ dynamics is then interpreted as part of the correlated two-pion spectral function.
Similarly, we include in the isoscalar spectral function the correlated $K\bar K$ and $\rho\pi$
continua, where the former also includes the $\phi$-meson pole.
The direct coupling terms remain essential for the transition to the asymptotic regime. This gives a
cleaner separation between long-range continuum dynamics, coherent vector-meson contributions, and direct
quark-current dynamics. With just 7 parameters, we can describe the large body on form-factor data and
differential cross sections from electron-proton scattering in the space-like region.
\end{abstract}

\maketitle


\section{Introduction}
\label{sec:introduction}

Electromagnetic form-factors are central observables for the structure of the nucleon. They describe how the electromagnetic current couples to the extended charge and magnetization distributions of the proton and neutron. In elastic electron--nucleon scattering, the form-factors enter the matrix element of the electromagnetic current and determine the measured cross sections. At low momentum transfer they are sensitive to the spatial distributions of charge and magnetization; at high momentum transfer they probe shorter-distance quark configurations. For reviews on the nucleon form-factors, see
\cite{PerdrisatPunjabiVanderhaeghen2007, Carlson:2015jba,Pacetti:2014jai,LinHammerMeissner2021,Gao:2021sml,Lin:2024rak}

Historically, vector-meson dominance (VMD) provided a successful phenomenological description of the low- and intermediate-\Qsq{} behavior~\cite{Sakurai1969,KrollLeeZumino1967,IachelloJacksonLande1973}. 
In this picture the virtual photon couples first to neutral vector mesons, most prominently the $\rho$, $\omega$, and $\phi$, which then couple to the nucleon. This gives form-factors with pole-like behavior and naturally accounts for the approximate dipole falloff observed at low and intermediate spacelike momentum transfer. Pure VMD, however, does not by itself reproduce the perturbative-QCD scaling behavior expected at large \Qsq{}. Furthermore,
it underestimates the nucleon electromagnetic radii.

The GK model was introduced to bridge these two descriptions. It retained the successful VMD description at moderate \Qsq{}, but modified the vector-meson amplitudes by intrinsic form-factors chosen to satisfy the perturbative-QCD scaling laws at large \Qsq{}. In this way the model connected meson-dominated dynamics with quark-current dynamics in one semiphenomenological framework~\cite{GariKruempelmann1984,GariKruempelmann1985,GariKruempelmann1986,GariKruempelmann1992}.
For other models based on VMD and extensions thereof, 
see e.g.~\cite{Lomon2001,Dubnicka:1992ih,Chen:2023oqs,Kuzmin:2024ozz}.

The purpose of the present work is to revisit the isovector part of the GK model in light of modern dispersive results for the two-pion continuum, see e.g.~\cite{Hoehler1976,Mergell:1995bf,HammerMeissner2004,BelushkinHammerMeissner2007,LinHammerMeissner2021,Lin:2021umk,Lin:2021xrc,Lin:2023fhr} and also the isoscalar part by including the $K\bar K$ and $\rho\pi$
continua~\cite{Hammer:1998rz,Hammer:1999uf,Meissner:1997qt}.
In the original GK model, the low-mass isovector strength was represented effectively by a $\rho$-meson term and by the direct coupling contribution. Modern dispersive analyses show that the physical low-mass isovector spectral function is not an isolated $\rho$ pole, but a correlated two-pion continuum with the $\rho$ resonance embedded in it,
as first shown in Refs.~\cite{FrazerFulco1959,FrazerFulco1960} and most recently updated in ~\cite{Hoferichter2016}. In particular, the enhancement on the left shoulder of the $\rho$ is very important to get the correct isovector radii. This was
first stressed in Ref.~\cite{Hohler:1974eq}. In fact, in the extended GK model of Ref.~\cite{Lomon2001},
the two-pion continuum from the dispersion
relation analysis was included in approximate analytic form from Ref~\cite{Mergell:1995bf}, but this is no longer state of
the art and we include it here directly without any approximation.

The main result is that the explicit two-pion continuum can replace the phenomenological $\rho$-pole contribution in the isovector Dirac form-factor.  While the $\rho$ pole term is subsituted by the two-pion contribution, the direct coupling terms remain, however,  essential, because they provide the nonresonant quark-current component needed for the transition to the perturbative-QCD regime.

Similarly, in the isoscalar channel, we follow the work of Ref.~\cite{BelushkinHammerMeissner2007} and include the
$K\bar K$ and $\pi\rho$ continua, drawing upon the work from Refs.~\cite{Hammer:1998rz,Hammer:1999uf} for the $K\bar K$
contribution and Ref.~\cite{Meissner:1997qt} for the $\pi\rho$ one. Note the $K\bar K$ continumm also includes the
$\phi$-meson pole and thus an explicit $\phi$ is no longer needed. Furthermore, there are strong cancellations
between these two contributions. This is of relevance since otherwise too large strange nucleon form factors
would emerge, see the discussions in Refs.~\cite{Jaffe:1989mj,Hammer:1995de,Meissner:1997qt}.

The manuscript is organized as follows: In Sect.~\ref{sec:gk-model} we collect the basic definitions
pertinent to the nucleon electromagnetic form-factors and discuss the original GK model, in 
particular also its strengths and limitations. The two-pion contribution as  well as the
$K\bar K$ and $\pi\rho$ contributions are discussed in Sect.~\ref{sec:two-meson-continuum} and included into the GK model.
In Sect.~\ref{sec:interpretation}, we discuss the physics of the
improved GK approach. Then, in Sect.~\ref{sec:fits-results}, we present our fits and show the resulting form-factors,
the form-factor ratios and the differential cross sections. We consider here two approaches. In the first one,
we just include the $2\pi$ continuum and the isoscalar spectral functions just contains the $\omega$. The second
and improved one also includes the isoscalar continua.
Finally, Sect.~\ref{sec:conclusions} contains a summary and the conclusions.

\section{Electromagnetic form-factors and the original GK model}
\label{sec:gk-model}

\subsection{Dirac, Pauli, and Sachs form-factors}
\label{subsec:dirac-pauli-sachs}

The electromagnetic current matrix element of the nucleon is written in terms of the Dirac and Pauli form-factors, $\Fone$ and $\Ftwo$, as
\begin{eqnarray}
&&\langle N(p')|J^{\mu}|N(p)\rangle \nonumber\\
&& = \bar u(p')\left[
 \gamma^{\mu}\Fone(\Qsq)
 + \frac{i\sigma^{\mu\nu}q_{\nu}}{2m_N}\Ftwo(\Qsq)
 \right]u(p),
\label{eq:current-matrix-element}
\end{eqnarray}
where $m_N$ is the nucleon mass, $q=p'-p$ and $\Qsq=-q^{2}>0$ in the spacelike region. The Sachs form-factors are
\begin{eqnarray}
  \GE(\Qsq) &=& \Fone(\Qsq)-\tau\Ftwo(\Qsq),\nonumber\\
  \GM(\Qsq) &=& \Fone(\Qsq)+\Ftwo(\Qsq),
\label{eq:sachs}
\end{eqnarray}
with
\begin{equation}
\tau=\frac{\Qsq}{4m_N^{2}}.
\label{eq:tau}
\end{equation}
The proton and neutron form-factors are decomposed into isoscalar and isovector
combinations according to 
\begin{equation}
F_{i}^{p,n}(\Qsq)=\frac{1}{2}\left[F_{i}^{S}(\Qsq)\pm F_{i}^{V}(\Qsq)\right],\qquad i=1,2,
\label{eq:pn-SV}
\end{equation}
where the upper sign refers to the proton and the lower sign to the neutron. Equivalently,
\begin{eqnarray}
F_{i}^{S}(\Qsq) &=& F_{i}^{p}(\Qsq)+F_{i}^{n}(\Qsq), \nonumber\\
F_{i}^{V}(\Qsq) &=& F_{i}^{p}(\Qsq)-F_{i}^{n}(\Qsq).
\label{eq:SV-def}
\end{eqnarray}
At $\Qsq=0$, the normalizations are
\begin{align}
F_{1}^{p}(0)&=1, & F_{1}^{n}(0)&=0, \\
F_{2}^{p}(0)&=\kappa_{p}, & F_{2}^{n}(0)&=\kappa_{n}~,
\label{eq:normalizations-pn}
\end{align}
with $\kappa_{p/n}$ the anomalous magnetic moment of the proton/neutron.
In the isoscalar/isovector basis this implies
\begin{eqnarray}
F_{1}^{S}(0)&=&F_{1}^{V}(0)=1, \\
F_{2}^{S}(0)&=&\kappa_{S}=\kappa_{p}+\kappa_{n}\simeq -0.120, \\
F_{2}^{V}(0)&=&\kappa_{V}=\kappa_{p}-\kappa_{n}\simeq 3.706.
\label{eq:normalizations-SV}
\end{eqnarray}
The Sachs form-factors, with the exception of $G_E^n(Q^2)$ are conventionally normalized to
the dipole form-factor $G_D(Q^2)$ given by 
\begin{equation}
G_D(Q^2) = \left(1 + \frac{Q^2}{0.71\,{\rm GeV}^2}\right)^{-2}~.
\end{equation}

\subsection{Vector-meson dominance and GK interpolation}
\label{subsec:vmd-interpolation}

VMD relates the hadronic electromagnetic current to low-lying neutral vector mesons. In the simplest form, one writes a form-factor as a sum of pole terms,
\begin{equation}
F_i(\Qsq)=\sum_V c_{iV}\frac{M_V^{2}}{M_V^{2}+\Qsq}~,
\label{eq:vmd-sum}
\end{equation}
where $M_V$ is the vector meson mass and the coefficients $c_{iV}$ encode effective vector-meson--nucleon couplings. In the isovector channel the dominant low-mass strength is associated with the $\rho$ channel, while the isoscalar channel contains the $\omega$ and $\phi(1020)$ contributions.

The GK model extends VMD by multiplying the pole contributions and the direct term by intrinsic form-factors. These factors suppress the hadronic amplitudes at high momentum transfer and enforce the perturbative-QCD power laws
\begin{equation}
\Fone(\Qsq)\sim Q^{-4},~
\Ftwo(\Qsq)\sim Q^{-6},~
\frac{\Ftwo(\Qsq)}{\Fone(\Qsq)}\sim Q^{-2},
\label{eq:pqcd-scaling}
\end{equation}
up to logarithmic corrections. A schematic representation of the intrinsic factors is
\begin{align}
F_{1}^{\mathrm{int}}(\Qsq)
  &=\left(\frac{\Lambda_{1}^{2}}{\Lambda_{1}^{2}+\Qsq}\right)^{2},
\label{eq:F1-int}\\
F_{2}^{\mathrm{int}}(\Qsq)
  &=\left(\frac{\Lambda_{1}^{2}}{\Lambda_{1}^{2}+\Qsq}\right)^{2}
    \left(\frac{\Lambda_{2}^{2}}{\Lambda_{2}^{2}+\Qsq}\right),
\label{eq:F2-int}
\end{align}
The precise implementation of the intrinsic factors follows the original GK convention in the numerical
analysis. The equations above display the structural form of the ansatz. Note that in the original GK model,
an additional
logarithmic term for large momentum transfers was included. As it has only a minor impact, it will be
neglected here.

\subsection{Baseline GK parametrization}
\label{subsec:baseline-parametrization}

With this convention, the isovector Dirac form-factor of the original GK ansatz may be written as
\begin{equation}
F_{1}^{V}(\Qsq)=\left[g_{\rho}\frac{M_\rho^2}{M_\rho^2+Q^2}+\left(1-g_{\rho}\right)\right]F_1^{\mathrm{int},V}(\Qsq),
\label{eq:baseline-F1V}
\end{equation}
with the corresponding Pauli form-factor contribution
\begin{eqnarray}
F_{2}^{V}(\Qsq)&=&\left[
 g_{\rho}\kappa_{\rho}\frac{M_{\rho}^{2}}{M_{\rho}^{2}+\Qsq}
 +\left(\kappa_V-g_{\rho}\kappa_{\rho}\right)
 \right]\nonumber \\
&\times& F_2^{\mathrm{int},V}(\Qsq).
\label{eq:baseline-F2V}
\end{eqnarray}
Here, the Pauli normalization is not factored out as an overall $\kappa_V$. Instead, the pole strength and the direct strength are fitted in the same normalization as used in the fit table. Since $F_2^{\mathrm{int},V}(0)=1$, Eq.~\eqref{eq:baseline-F2V} gives $F_2^V(0)=\kappa_V=3.706$ directly. The effective $\rho$ term is proportional to $g_{\rho}\kappa_{\rho}$, while the direct Pauli strength is $3.706-g_{\rho}\kappa_{\rho}$.

The isoscalar channel contains the dominant $\omega$ and $\phi(1020)$ pole contributions and the direct term.
For the Dirac form-factor this amounts to
\begin{eqnarray}
F_{1}^{S}(\Qsq)&=&\Biggl[
 g_{\omega}\frac{M_{\omega}^{2}}{M_{\omega}^{2}+\Qsq} 
 +  g_{\phi}\frac{M_{\phi}^{2}}{M_{\phi}^{2}+\Qsq}\nonumber\\
 &+&\left(1-g_{\omega}-g_{\phi}\right)
 \Biggr]\,F_{1}^{\mathrm{int},S}(\Qsq),
\label{eq:isoscalar-baseline-F1}
\end{eqnarray}
where the bracket is normalized to $F_1^S(0)\kappa_S=1$. In the Pauli isoscalar channel the same convention as in the fit table is used: the overall anomalous moment is not pulled out. The minimal form used here is
\begin{eqnarray}
F_{2}^{S}(\Qsq)&=&\left[
 g_{\omega}\kappa_{\omega}\frac{M_{\omega}^{2}}{M_{\omega}^{2}+\Qsq}
 +\left(\kappa_S-g_{\omega}\kappa_{\omega}\right)
 \right]\nonumber\\
&\times& F_{2}^{\mathrm{int},S}(\Qsq).
\label{eq:isoscalar-baseline-F2}
\end{eqnarray}
This gives $F_2^S(0)=\kappa_S=-0.120$ directly. The fitted pole strength is $g_{\omega}\kappa_{\omega}$ and the corresponding direct Pauli strength is $-0.120-g_{\omega}\kappa_{\omega}$.
The $\phi$ pole denotes the ordinary $\phi(1020)$, not a separate higher $\phi'$ resonance. Its role in the GK framework is tied to possible strange isoscalar strength. In the final Gari--Kr\"umpelmann analysis, the electric neutron form-factor was reanalyzed with special attention to this strange-quark component. The result was that the $\phi$-associated strange contribution can modify $G_E^n$ at low momentum transfer without changing the asymptotic perturbative-QCD behavior.
In more recent approaches, much of this $\phi(1020)$ contribution is generated by the $\bar{K}K$ continuum
and this is largely  cancelled by the $\pi\rho$ exchange, see e.g.
Ref.~\cite{Meissner:1997qt}. In what follows, we will consider a minimal setup just including the
$\omega$ meson.

\subsection{Direct coupling}
\label{subsec:direct-coupling}

The GK ansatz also contains a direct, nonresonant coupling of the electromagnetic current to the nucleon. This term is not an arbitrary background. It represents the part of the photon-induced quark--antiquark amplitude which is not projected onto a coherent intermediate vector-meson state.

\begin{figure}[htbp]
\centering
\includegraphics[width=0.6\linewidth]{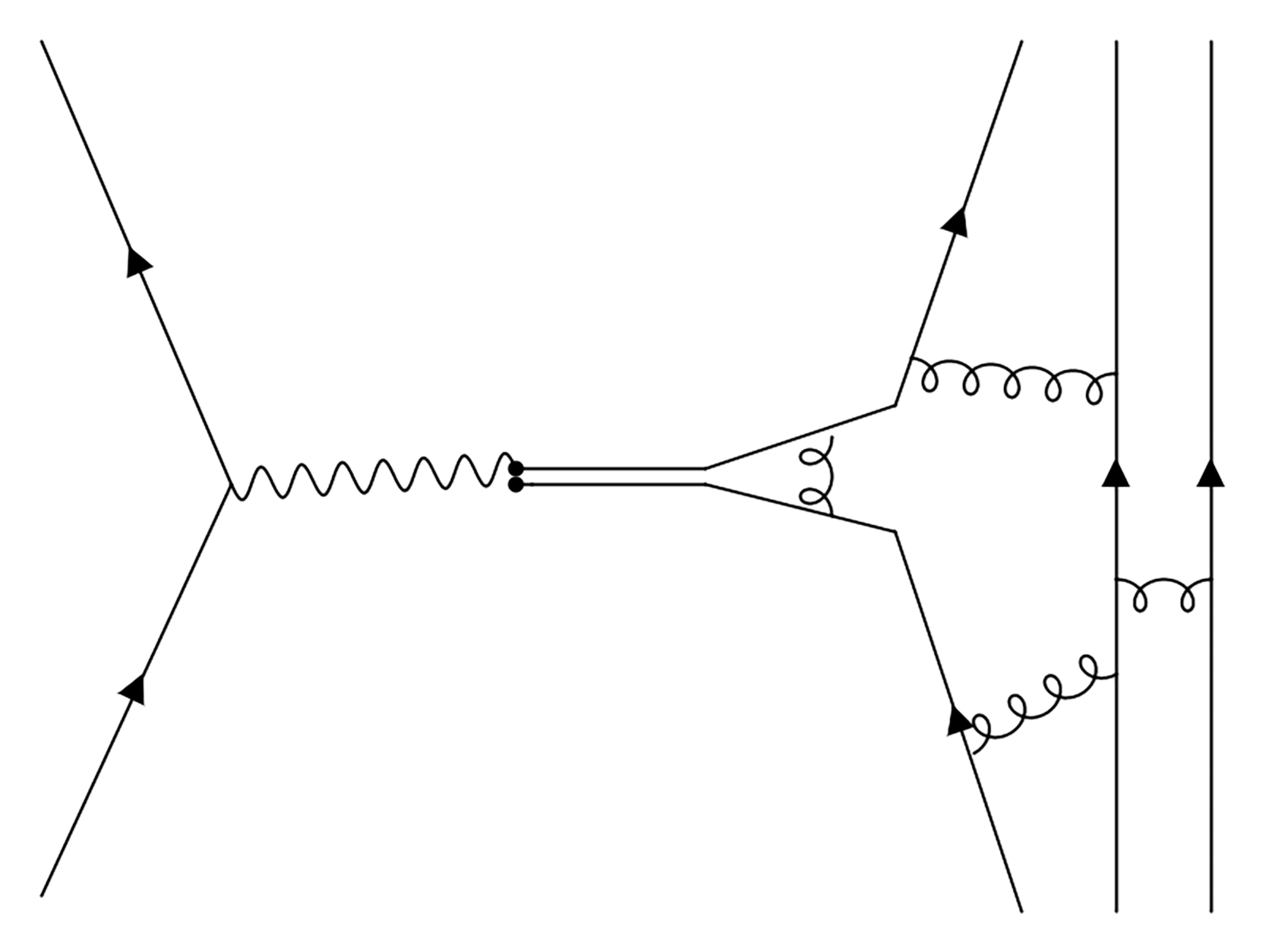}
\caption{Schematic interpretation of the direct coupling contribution. The photon-induced $\qqbar$ configuration is not projected onto an intermediate vector-meson pole, but is resolved and coupled directly to the valence-quark structure of the nucleon.}
\label{fig:direct-coupling}
\end{figure}

In ordinary VMD, the virtual photon first converts into a neutral vector meson, and this meson then couples to the nucleon. The corresponding term has the pole structure characteristic of the vector-meson propagator. The direct term describes a different mechanism. The photon may still create a $\qqbar$ configuration with the appropriate quantum numbers, but this configuration need not remain bound as an intermediate meson. Instead, it can be resolved into quark degrees of freedom and couple directly to the valence-quark structure of the nucleon, see Fig.~\ref{fig:direct-coupling}.

This interpretation gives the direct term a natural role in the GK framework. At low and intermediate momentum transfer, the electromagnetic current is dominated by coherent hadronic degrees of freedom and by the VMD pole terms. With increasing \Qsq{}, however, the resolving power of the virtual photon increases. The electromagnetic probe becomes sensitive to shorter-distance quark configurations, and the nonresonant direct coupling becomes increasingly relevant. Together with the intrinsic form-factors, the direct terms connect the VMD region to the perturbative-QCD regime.

\subsection{Strengths and limitations of the original model}
\label{subsec:strengths-limitations}

The original GK model was one of the first semiphenomenological descriptions of the electromagnetic nucleon form-factors that attempted to cover the complete available range of momentum transfer in a unified framework. It was not restricted to either the low-\Qsq{} VMD region or the asymptotic perturbative-QCD region. Instead, it provided a continuous interpolation between both domains and treated proton and neutron form-factors within the same isoscalar/isovector structure.

A central strength of the model was that it was not merely a numerical fit. The parametrization was tied to a physical interpretation of the underlying processes. The vector-meson pole terms represented the hadronic part of the electromagnetic current, where the virtual photon couples through intermediate $\rho$, $\omega$, and $\phi$ mesons. The pole denominators described the photon--meson part of the process, while the residues encoded the effective meson--nucleon couplings. The intrinsic factors described the transition to short-distance quark dynamics and enforced the perturbative-QCD scaling behavior at large \Qsq{}.

The most visible limitation was the role of the cutoff parameters $\Lambda_i$. These parameters were introduced as intrinsic mass scales controlling the transition from meson-dominated to quark-dominated behavior. However, their fitted numerical values were close to typical hadronic and vector-meson mass scales. Modern fits of the original minimal form tend to require values of approximately
\begin{equation}
\Lambda_1\simeq 1.4\GeV,\qquad
\Lambda_2\simeq 0.9\GeV.
\label{eq:typical-lambdas}
\end{equation}
The first scale lies already above the light vector-meson region, while the second remains close to typical vector-meson masses. This weakens the clean separation between vector-meson pole dynamics and intrinsic short-distance structure. It suggests that missing spectral and continuum physics is partly absorbed into the cutoff parameters rather than represented explicitly.

A second limitation was the simplified treatment of the isovector spectral function. In the original parametrization, the isovector channel was represented mainly by an effective $\rho$ contribution. This captures the dominant resonance scale, but it does not explicitly include the two-pion continuum. Missing $\twopi$ strength was therefore absorbed into pole residues and intrinsic cutoff parameters. The present work addresses this limitation directly. In addition, we improve the isoscalar
spectral function, as in the original GK model only the $\omega$ and $\phi$ poles were considered. 

\subsection{The two-meson continua}
\label{sec:two-meson-continuum}

In this section, we discuss first the isovector  two-pion continuum and its inclusion in the GK model.
Then, we turn to the isoscalar $K\bar K$ and $\pi\rho$ continua and their addition.

\subsubsection{The two-pion continuum from a dispersive approach}
\label{sec:two-pion-continuum}


The lowest hadronic intermediate state in the isovector electromagnetic channel is the two-pion state.
The branch cut of the isovector form-factors begins at
\begin{equation}
t_\pi = 4M_{\pi}^{2},
\label{eq:threshold}
\end{equation}
where $t$ is the invariant momentum transfer, $t=-\Qsq$ in the spacelike region and $M_\pi$ is the charged pion mass.
A schematic dispersion relation for the isovector form-factors is
\begin{equation}
F_i^V(t)=\frac{1}{\pi}\int_{t_\pi}^{\infty}
\frac{\operatorname{Im}F_i^V(t')}{t'-t-i0}\,dt',\qquad i=1,2.
\label{eq:dispersion-t}
\end{equation}
In the spacelike region this becomes
\begin{equation}
F_i^V(-\Qsq)=\frac{1}{\pi}\int_{t_\pi}^{\infty}
\frac{\operatorname{Im}F_i^V(t')}{t'+\Qsq}\,dt'.
\label{eq:dispersion-Q2}
\end{equation}
The two-pion contribution is constrained by unitarity. In the elastic region, the isovector Sachs spectral
functions are related to the pion electromagnetic form-factor and to the $P$-wave amplitudes for
$\pi\pi\to\bar NN$, denoted as $f_\pm^{1}(t)$. This gives
\begin{align}
\operatorname{Im}G_E^V(t)&= \frac{q_{\pi}^{3}(t)}{m_N\sqrt{t}}F_{\pi}^{V*}(t)f_{+}^{1}(t),
\label{eq:ImGEV}\\
\operatorname{Im}G_M^V(t)&= \frac{q_{\pi}^{3}(t)}{\sqrt{2t}}F_{\pi}^{V*}(t)f_{-}^{1}(t),
\label{eq:ImGMV}
\end{align}
The pion-nucleon amplitudes are taken from the Roy-Steiner analysis~\cite{Hoferichter:2015hva}.
Also, $F_{\pi}^{V}(t)$ denotes the electro\-magnetic vector form-factor of the charged pion, defined by
\begin{equation}
\langle \pi^{+}(p')|j_{\mathrm{em}}^{\mu}|\pi^{+}(p)\rangle
=(p+p')^{\mu}F_{\pi}^{V}(t),
~~ t=(p'-p)^2 .
\label{eq:pion-ff-definition}
\end{equation}
Elastic $\pi\pi$ intermediate states imply the unitarity condition
\begin{equation}
\operatorname{Im}F_{\pi}^{V}(t)=
\sin\delta_{1}^{1}(t)\,e^{-i\delta_{1}^{1}(t)}F_{\pi}^{V}(t)\,
\theta(t-t_{\pi})~,
\label{eq:pion-ff-unitarity}
\end{equation}
where $\delta_{1}^{1}(t)$ is the isovector $P$-wave $\pi\pi$ phase shift. This form of the unitarity
relation expresses Watson's final-state theorem: below significant inelastic thresholds the phase of $F_{\pi}^{V}$ is fixed by the elastic $\pi\pi$ scattering phase. If higher intermediate states are neglected, the solution can be written as a polynomial times the Omn\`es function,
\begin{eqnarray}
F_{\pi}^{V}(t) &=&P(t)\Omega_{1}^{1}(t),
\nonumber\\
\Omega_{1}^{1}(t)&=&\exp\left\{
\frac{t}{\pi}\int_{t_{\pi}}^{\infty}
\frac{\delta_{1}^{1}(t')}{t'(t'-t-i0)}\,dt'
\right\}.
\label{eq:omnes-representation}
\end{eqnarray}
This representation is a compact way of incorporating the correlated two-pion dynamics, including the broad $\rho$ enhancement, through the empirical $P$-wave phase shift. The remaining slowly varying part is contained in the polynomial $P(t)$. In practice, the description of the pion form-factor below the $1\,\mathrm{GeV}$ region is improved by including the isospin-violating $\rho$--$\omega$ mixing term. Following the dispersive representation used in Ref.~\cite{Hoferichter2016},
the pion vector form-factor is parameterized as
\begin{equation}
F_{\pi}^{V}(t)=\left(
1+\alpha t+
\frac{\epsilon_{\rho\omega} t}{M_{\omega}^{2}-iM_{\omega}\Gamma_{\omega}-t}
\right)\Omega_{1}^{1}(t).
\label{eq:pion-vector-form-factor}
\end{equation}
Here, $M_{\omega}$ and $\Gamma_{\omega}$ are the $\omega$ mass and width, while $\alpha$ and $\epsilon_{\rho\omega}$ are determined from fits to timelike pion-form-factor data. In particuar, $\epsilon_{\rho\omega}$ parameterizes the isospin-violating
$\rho$-$\omega$ mixing. We take the values from Ref.~\cite{Hoferichter2016} in what follows.
 The phase shifts used in the Omn\`es function are taken from Roy- and Roy-like-equation analyses; the spread between alternative phase-shift inputs and data sets gives part of the uncertainty in the dispersive spectral functions. Further, the pion momentum is
\begin{equation}
q_{\pi}(t)=\sqrt{\frac{t}{4}-M_{\pi}^{2}}.
\label{eq:qpi}
\end{equation}
The factor $q_{\pi}^{3}(t)$ reflects the $P$-wave nature of the two-pion state and gives the correct threshold behavior. The $\rho$ resonance appears dynamically as the dominant enhancement in the pion form-factor, but the full spectral function is not identical to a narrow $\rho$ pole. It includes threshold behavior, finite-width effects, the low-mass shoulder below the $\rho$ peak, and the correlated pion dynamics in the isovector channel.

\subsubsection{The isoscalar continuum}
\label{sec:isoscalar-continuum}

First, we consider the $K\bar K$ contribution. This has been worked out in detail in Refs.~\cite{Hammer:1998rz,Hammer:1999uf}.
It should be noted that the extraction is less precise than in the $2\pi$ isovector case, partly because there are less
precise data e.g. on kaon-nucleon scattering and the analytic continuation of the scattering amplitudes is more subtle.
The striking feature in the spectral function is a clear  $\phi$ resonance structure just above the $K\bar{K}$ threshold.
Also, there is no strong enhancement on the  left wing of the $\phi$ resonance which sits directly at the 
$K\bar{K}$ threshold. The resulting contribution to the nucleon form factors can be 
parameterized by a pole term at the $\phi$ mass:
\begin{eqnarray}
F_i^{(s,K\bar{K})}(t) &=&
\frac{1}{\pi} \int_{4M_K^2}^\infty
\frac{{\rm Im} \, F_i^{(s,K\bar{K})}(t')}{t'-t}dt' \nonumber\\
&\approx&
\frac{a_i^{K\bar{K}}}{M_\phi^2-t}\,,
\qquad i=1,2\,,
\label{KKcont}
\end{eqnarray}
with $a_1^{K\bar{K}}=0.1054\,$GeV$^2$ and $a_2^{K\bar{K}}=0.2284\,$GeV$^2$. This can be conveniently
included into the GK model as described in Sect.~\ref{subsec:incorporation}.

Similarly, the $\rho\pi$ contribution to the isoscalar spectral function was
calculated in Ref.~\cite{Meissner:1997qt}. 
The contribution of the $\rho\pi$ continuum can be evaluated
in terms of a dispersion integral which in turn can be represented
by an effective pole term for a fictitious $\omega'$ meson with 
a mass $M_{\omega'} = 1.12\,$GeV:
\begin{eqnarray}
F_i^{(s,\rho\pi)}(t) &=&
\frac{1}{\pi} \int_{(M_\pi +M_\rho)^2}^\infty
\frac{{\rm Im} \, F_i^{(s,\rho\pi)}(t')}{t'-t}dt' \nonumber\\
&\approx& 
\frac{a_i^{\rho\pi}}{M_{\omega'}^2-t}\,, \qquad i=1,2
\end{eqnarray}
with $a_1^{\rho\pi}=-1.01$ GeV$^2$ and $a_2^{\rho\pi}=-0.04$ GeV$^2$. This can also easily be
included in the GK model.

\subsubsection{Incorporation into the GK ansatz}
\label{subsec:incorporation}

The extended isovector Dirac form-factor is written in normalized schematic form as
\begin{equation}
F_1^V(\Qsq)=\left[
 g_{\twopi,1}\widehat F_{\twopi,1}^{\mathrm{disp}}(\Qsq)
 +\left(1-g_{\twopi,1}\right) \right]F_1^{\mathrm{int},V}(\Qsq).
\label{eq:extended-F1V}
\end{equation}
The corresponding Pauli form-factor is correspondingly written in the same convention:
\begin{eqnarray}
F_2^V(\Qsq)&=&\left[
 g_{\twopi,2}\widehat F_{\twopi,2}^{\mathrm{disp}}(\Qsq)
 +\left(3.706-g_{\twopi,2}\right)
 \right] \nonumber\\
&\times& F_2^{\mathrm{int},V}(\Qsq).
\label{eq:extended-F2V}
\end{eqnarray}
The bracketed structure in Eq.~\eqref{eq:extended-F1V} is normalized to unity at $\Qsq=0$. In contrast, Eq.~\eqref{eq:extended-F2V} is normalized directly to the isovector anomalous moment. Since $\widehat F_{\twopi,2}^{\mathrm{disp}}(0)=F_2^{\mathrm{int},V}(0)=1$, it gives $F_2^V(0)=3.706$ without introducing an external overall factor. The direct Pauli strength in the isovector channel is therefore $3.706-g_{\twopi,2}$. Analogously, the isoscalar Pauli expression is normalized directly to $F_2^S(0)=-0.120$.

The interpretation is direct. The coefficients $g_{\twopi,1}$ and $g_{\twopi,2}$ measure the strength assigned to the explicit low-mass two-pion continuum. The direct term has strength
\begin{equation}
g_{\mathrm{dir}}^{(1),V}=1-g_{\twopi,1}
\label{eq:gdir}
\end{equation}
in the Dirac channel. In the Pauli isovector channel the corresponding direct strength is
\begin{equation}
g_{\mathrm{dir}}^{(2),V}=3.706- g_{\twopi,2}~.
\label{eq:gdir-F2V}
\end{equation}
Both direct terms represent the nonresonant quark-current component.

The isoscalar Dirac form factor including the $K\bar K$ and $\rho\pi$ continua takes the form:
\begin{eqnarray}
F_{1}^{S}(Q^2)&=&\Bigg[g_{\omega}^1\frac{M_{\omega}^{2}}{M_{\omega}^{2}+Q^2} +  g_{\phi}^1\left[F_{\pi\rho,1}(Q^2)
+F_{K\bar{K},1}(Q^2)\right] \nonumber\\
&+&\left(1-g_{\omega}^1-\tilde{g}_{\phi}^1\right)
 \Bigg]\,F_{1}^{\mathrm{int}}(Q^2)~,
\end{eqnarray}
and similarly for the isoscalar Pauli form factor:
\begin{eqnarray}
F_{2}^{S}(Q^2)&=&\Bigg[g_{\omega}^2\frac{M_{\omega}^{2}}{M_{\omega}^{2}+Q^2}
+g_{\phi}^2\left[F_{\pi\rho,2}(Q^2)+F_{K\bar{K},2}(Q^2)\right] \nonumber\\
&+&\left(-0.12-g_{\omega}^2-\tilde{g}_{\phi}^2\right)~,
\Bigg] F_{2}^{\mathrm{int}}(Q^2
\end{eqnarray}
with
\begin{equation}
\tilde{g}_{\phi}^i\equiv {g}_{\phi}^i\left[F_{\pi\rho,i}(0)+F_{K\bar{K},i}(0)\right]~.
\end{equation}  
The interpretation is as direct as in the isoscalar case.

\section{Physical interpretation of the isovector and isoscalar channels}
\label{sec:interpretation}

The electromagnetic form-factors contain both isovector and isoscalar contributions. In the GK framework these contributions are interpreted as phenomenological representations of distinct physical mechanisms. Pole terms describe coherent vector-meson intermediate states, continuum terms describe multi-pion spectral strength, and direct terms represent nonresonant quark-current contributions not projected onto an intermediate meson pole.

In a general spectral representation, the isovector Dirac form-factor may receive contributions from
\begin{equation}
F_1^V=F_1^{\twopi}+F_1^{\rho'}+F_1^{\rho''}+F_1^{\mathrm{dir}}.
\label{eq:isovector-content}
\end{equation}
The original GK model used a more economical representation consisting of the dominant $\rho$ contribution and the direct term. A $\rho^\prime$ contribution was investigated in early fits, but its  contribution was found to be very small and it did not significantly improve the description of the data available at that time. From the present perspective this is natural: a substantial part of the low-mass isovector strength attributed effectively to the $\rho$ pole in the original model is actually the correlated two-pion continuum. Still, in refinements
of the model presented here such higher mass contributions might be considered.

The isoscalar channel has a different spectral content. In schematic form,
\begin{equation}
F_i^S=F_i^{\threepi}+F_i^{\omega}+F_i^{\phi}+F_i^{\omega'}+F_i^{\phi'}+F_i^{\mathrm{dir}},\qquad i=1,2.
\label{eq:isoscalar-content}
\end{equation}
The original GK model retained the dominant low-mass isoscalar pole contributions, $\omega$ and $\phi(1020)$, together with the direct term. Higher isoscalar poles such as $\omega'$ and $\phi'$ were investigated, but their contributions were negligible for the data available at the time. This is plausible because of their higher masses and because their residual effects can be absorbed into the effective pole residues and the direct term unless demanded by the data.

The three-pion continuum is the isoscalar analogue of the two-pion continuum. In principle it forms part of the full isoscalar spectral function, with the $\omega$ resonance embedded in correlated $\threepi$ dynamics. It was shown in Ref.~\cite{Bernard:1996cc} that although there is a anomalous threshold at $t_c = 8.9M_\pi^2$ close to the
three-pion cut starting at $t_0=9M_\pi^2$, it does not lead to any enhancement on the left shoulder of the $\omega$ due to phase space suppression. Thus, no explicit $F_i^{\threepi}$ is required.

The $\phi(1020)$ contribution plays a special role. Although the $\phi$ is dominantly an $s\bar s$ state in the ideal mixing limit, it can generate significant variations in $G_E^n$. This matters because $G_E^n$ is not measured as a free-neutron elastic observable. It is extracted from reactions on light nuclei, especially deuteron breakup and related quasielastic processes. These extractions depend on the nuclear wave function, final-state interactions, meson-exchange currents, and the nucleon--nucleon potential. A consistent analysis should therefore use potential models and electromagnetic form-factors based on compatible meson--nucleon dynamics. For discussion, we refer to the published meson-exchange study 
in Ref.~\cite{GariDeisterKruempelmannMahlke1991}.  Also, as noted before, for a truely consistent inclusion of the
$\phi$ as generated from the $\bar{K}K$ continuum, one needs to also include the $\rho\pi$ continuum of opposite
sign, as it is done here.

\section{Fits and results}
\label{sec:fits-results}

We will consider two scenarios. In the first one, we have the two-pion continuum in the isovector and the
$\omega$-meson in the isoscalar spectral function. This leaves us with 6 fit parameters. In the second, and physically
more realistic scenario, we add the $K\bar K$ and $\rho\pi$ continua. In that approach, we deal with 8 fit parameters.
However, as we will see, all fits are equally well described with $\Lambda_1 = \Lambda_2$, so we effectively have
5 and 7 parameters only.

\subsection{Fit strategy}
\label{subsec:fit-strategy}

First, as a sanity check, we performed fits with an explicit pole and the two-pion continuum. These fits are not 
further discussed as it was found that an explicit  $\rho$ is not needed any more if the two-pion continuum constructed
by dispersive methods as described above is included. Second, we fit to the differential cross sections from
MAMI and Jefferson Lab,  the  form-factor ratios $\mu_P G_E^p(Q^2)/G_M^p(Q^2)$ and $\mu_n G_E^n(Q^2)/G_M^n(Q^2)$
from Jefferson Lab and Mainz as well as the
world data on the proton and neutron electric and magnetic form-factors.  The data basis is a combination
of Refs.~\cite{GariKruempelmann1985,Lin:2021xrc} with the addition of the data from
Refs.~\cite{Riordan:2010id,Schlimme:2013eoz}.
Note that we do not prune this data basis, that is,
some of the data sets are not consistent.
Alltogether, we have just six parameters for the first scenario,
\begin{equation}
g_{2\pi,1}~,~~ g_{2\pi,2}~,~~g_{\omega,1}~,~~g_{\omega,2}~,~~\Lambda_1~,~~\Lambda_2~,
\end{equation}  
and eight parameters for the sceond one:
\begin{equation}
g_{2\pi,1}~,~~ g_{2\pi,2}~,~~g_{\omega,1}~,~~g_{\omega,2}~,~~g_{\phi,1}~,~~g_{\phi,2}~,~~\Lambda_1~,~~\Lambda_2~,
\end{equation}  
with the notation $g_{\omega,1}= g_\omega$, $g_{\omega,2}=\kappa_\omega g_\omega$, $g_{\phi,1}= g_\phi$ and $g_{\phi,2}=\kappa_\phi g_\phi$
These are determined from a least $\chi^2$ to the one-sigma error bars of the data. We minimize the weighted chi-squared function
\begin{align}
	\chi^2
	=
	\chi^2_{\rm PRad}
	+
	\chi^2_{\rm MAMI}
	+
	\sum_{a=1}^{6} w_a\,\chi^2_{F_a},
\end{align}
where, for a given data set $D$,
\begin{align}
	\chi^2_D
	=
	\sum_{i\in D}
	\frac{
		\left(y_i^{\rm th}-y_i^{\rm exp}\right)^2
	}{
		\left(\delta y_i\right)^2
	},
\end{align}
with
\begin{align}
	\left(\delta y_i\right)^2
	=
	\left(\delta y_i^{\rm stat}\right)^2
	+
	\left(\delta y_i^{\rm syst}\right)^2~,.
\end{align}
where $\delta  y_i^{\rm stat}$ and $\delta y_i^{\rm syst}$ are the corresponding statistcal and systematic uncertainties
of the data. Further, $F_a$ denotes the six form-factor (ratio) data sets and $w_a$ the corresponding weights.
Assigning a larger weight to some data sets is done to allow for a better
description of the form-factors. We provide a measure of the uncertainty as discussed in Sect.~\ref{sec:unc}.

However, in these six parameter fits, we find that $\Lambda_1 \simeq \Lambda_2$, so in the minimal
approach, which we consider in what follows, we set $\Lambda_2 = \Lambda_1 =\Lambda$. We observe that these
five parameter fits have the same $\chi^2/{\rm dof}$ as the ones with six parameters.
The fit algorithm is a modified deepest access algorithm, which uses a polynom approach in the
deepest valley. The start values were the old values from the GK formfactors. 
An independent implementation using the Minuit minimization algorithm~\cite{Minuit:cern} through its Julia interface~\cite{Minuit:julia} yields consistent results.

\subsection{Uncertainties}
\label{sec:unc}

In the present work, we only consider the statistical uncertainties of the model parameters,
which are propagated from the errors of the experimental data points included in the fits. Since
the physical picture underlying our extended GK parametrization is relatively straightforward and
the model contains only five free parameters, we do not introduce additional model uncertainties
at this stage.

To estimate the statistical uncertainties, we perform a bootstrap analysis by generating 5000
sets of pseudo-data samples based on the fitted data points. For each data point, the pseudo-data
are generated according to an uncorrelated Gaussian distribution, with the corresponding
experimental uncertainty taken as the standard deviation. By independently fitting these
bootstrap pseudo-data samples, we obtain the probability distributions of the model parameters
and subsequently determine their statistical uncertainties.
For details, we refer to~\cite{LinHammerMeissner2021}.

\subsection{Fits to the MAMI cross section data}
\label{sec:mami}

\begin{figure}[b]
\centering
\includegraphics[width=\linewidth]{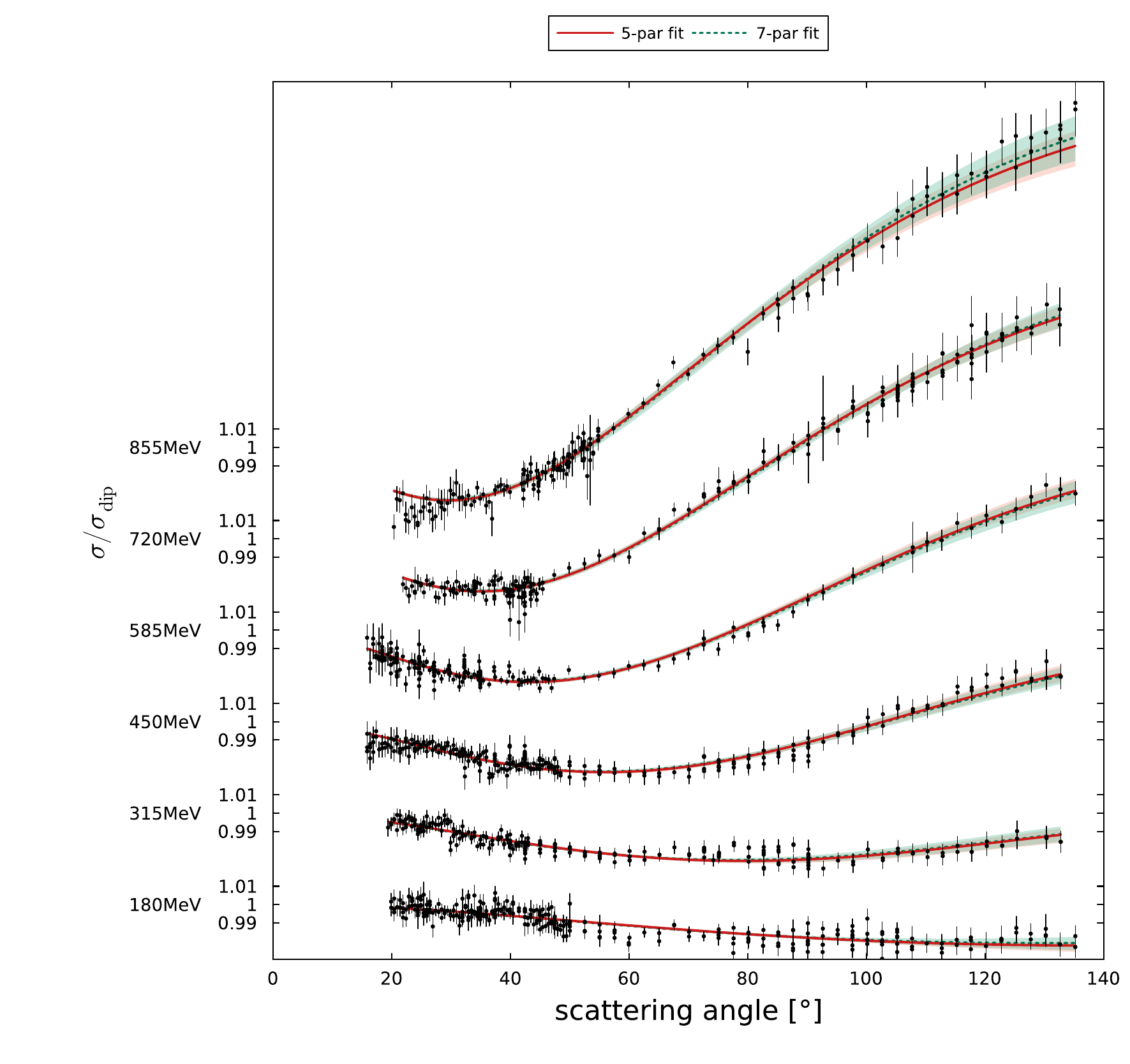}
\caption{Fitted cross sections for the best fit (solid lines) with uncertainties (bands)
  to the  data from MAMI~\cite{A1:2013fsc}
  normalized to the dipole cross section~$\sigma_{\rm dip}$.
  Red/green color refers to scenario~1/2.}
\label{fig:cross-sectionsMonly}
\end{figure}

As a first test of the model, we fit to the 1422 precision data points from the A1 collaboration~\cite{A1:2013fsc}.
This is entirely new for this type of model, which so far has only been used to fit to form-factor data.
In these fits, we neglect two-photon exchange corrections.  The resulting best fits are shown in
Fig.~\ref{fig:cross-sectionsMonly} for both scenarios, with  a $\chi^2/{\rm dof} = 0.96$ and $0.95$, respectively.
The fit parameters are collected in Tab.~\ref{tab:mami}.
\begin{table}[t]
\begin{center}
\begin{tabular}{|l|cc|}
\hline
Parameter & Scenario~1 & Scenario~2 \\
\hline
$g_{2\pi,1}$       & 0.694(178)      & 0.728(256)\\
$g_{2\pi,2} $      &  5.484(1701)   & 3.575(1273)\\
$g_{\omega,1}$ & 1.053(405)    & 0.572(843) \\
$g_{\omega,2} $ & -2.465(1021)  & 0.373(1668)\\
$g_{\phi,1}$        &  ---                &  -0.345(1000) \\
$g_{\phi,2}$        &  ---                &  -10.0(2.3)\\
$\Lambda$ [GeV] & 1.691(732) &  1.444(586) \\
\hline
\end{tabular}
\end{center}
\caption{Values of the fit parameters for the two scenarios. See the text for more details.}
\label{tab:mami}
\end{table}
Even though  we did not apply any (floating) normalization factors
as done in the original analysis and also in the dispersion-theoretical approach,
one finds a rather good description of these data. This is quite a strong benchmark
for this rather simple model.
The resulting proton radii are 
\begin{eqnarray}
{\rm Scenario}~1:&& ~~ r_E^p = 0.844(2)\,{\rm fm }, ~~~ r_M^p = 0.855(7)\,{\rm fm},\nonumber\\
{\rm Scenario}~1: &&~~ r_E^p = 0.846(3)\,{\rm fm }, ~~~ r_M^p = 0.848(10)\,{\rm fm},\nonumber\\
&&
\end{eqnarray}
Both these radii are consistent with the values found in dispersion theory, cf. Ref.~\cite{Lorenz:2012tm} for
a first discussion of the A1 data in view of the dispersive approach with $r_E^p = 0.84(1)\,$fm and
$r_M^p = 0.86(3)\,$fm. Again, the magnetic radius comes out  sizeably  larger than found in the
original A1 analysis. Note also that with the parameters given in Tab.~\ref{tab:mami}, the PRad
data~\cite{Xiong:2019umf} are also described. In the next step, we will now perform fits on
all cross section and form-factor data, in the spirit of the original GK model.

\subsection{Fits to form-factors and cross sections}
\label{subsec:gmn-cross-sections}

The best fits with a $\chi^2/{\rm dof} = 2.65$ and $2.49$ are obtained for the  values of the
parameters displayed in Tab.~\ref{tab:global}.
\begin{table}[t]
\begin{center}
\begin{tabular}{|l|cc|}
\hline
Parameter & Scenario~1 & Scenario~2 \\
\hline
$g_{2\pi,1}$       & 0.606(27)      & 0.550(30)\\
$g_{2\pi,2} $      &  2.936(139)   & 2.694(144)\\
$g_{\omega,1}$ & 0.963(198)    & 1.022(27) \\
$g_{\omega,2} $ & -0.121(22)  & -0.077(178)\\
$g_{\phi,1}$        &  ---              &  -0.174(61) \\
$g_{\phi,2}$        &  ---              &  -0.197(1132)\\
$\Lambda$ [GeV] & 1.494(39) &  1.430(34) \\
\hline
\end{tabular}
\end{center}
\caption{Values of the fit parameters for the two scenarios. See the text for more details.}
\label{tab:global}
\end{table}

\begin{figure*}[htbp]
\includegraphics[width=0.90\linewidth]{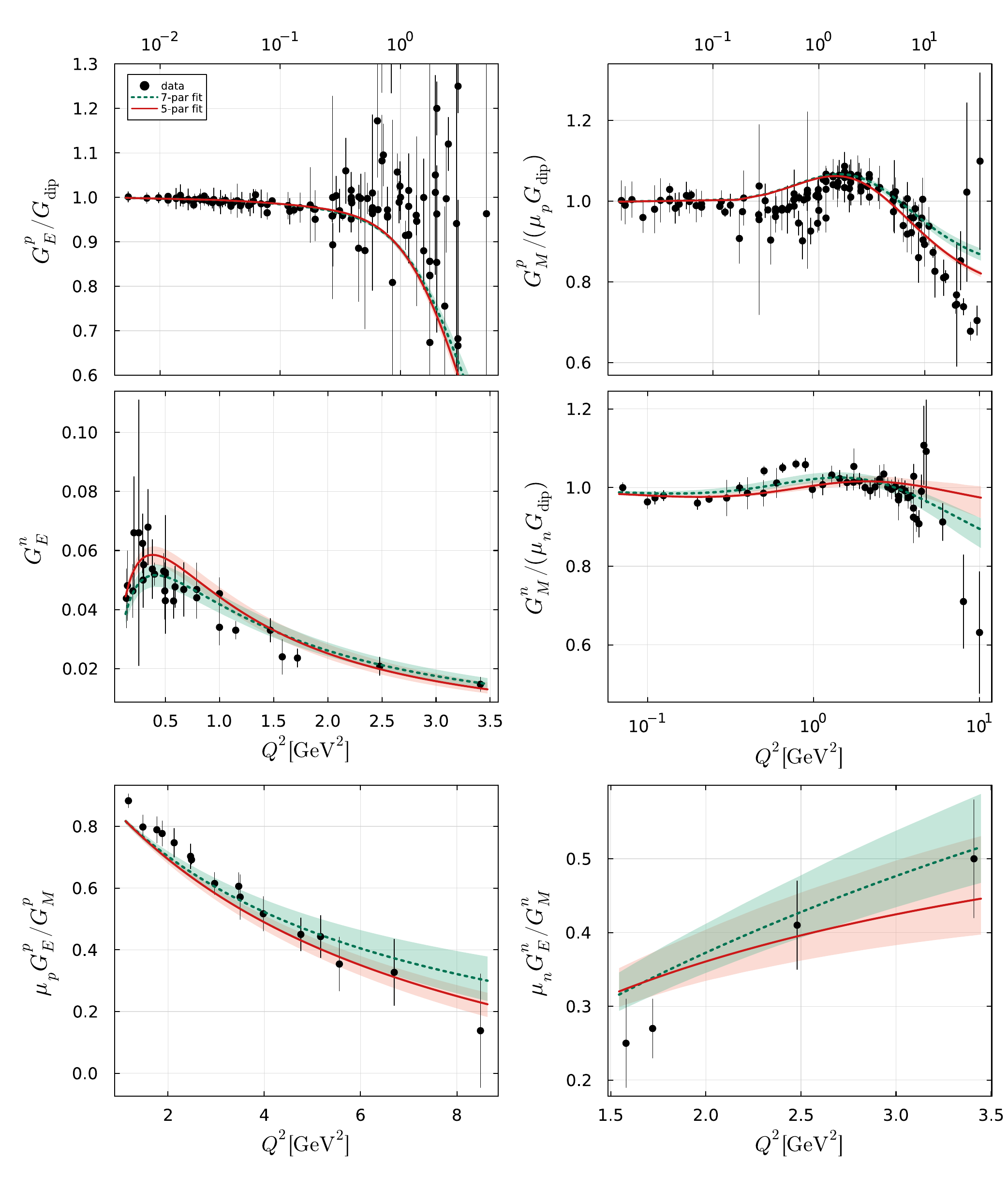} 
\caption{Best fit for the form-factors and the form-factor ratios (solid red lines) in comparison with the data.
  The uncertainties are displayed by the red bands. The dashed blue lines show the best fit with equal weights
  and the dashed-dotted balck lines the fit without $G_{E,M}^p(Q^2)$ and weights as in the best fit. 
  Red/green color refers to scenario~1/2. See the text for more discussions.}
\label{fig:ffs}
\end{figure*}

\begin{figure}[htbp]
\centering
\includegraphics[width=0.99\linewidth]{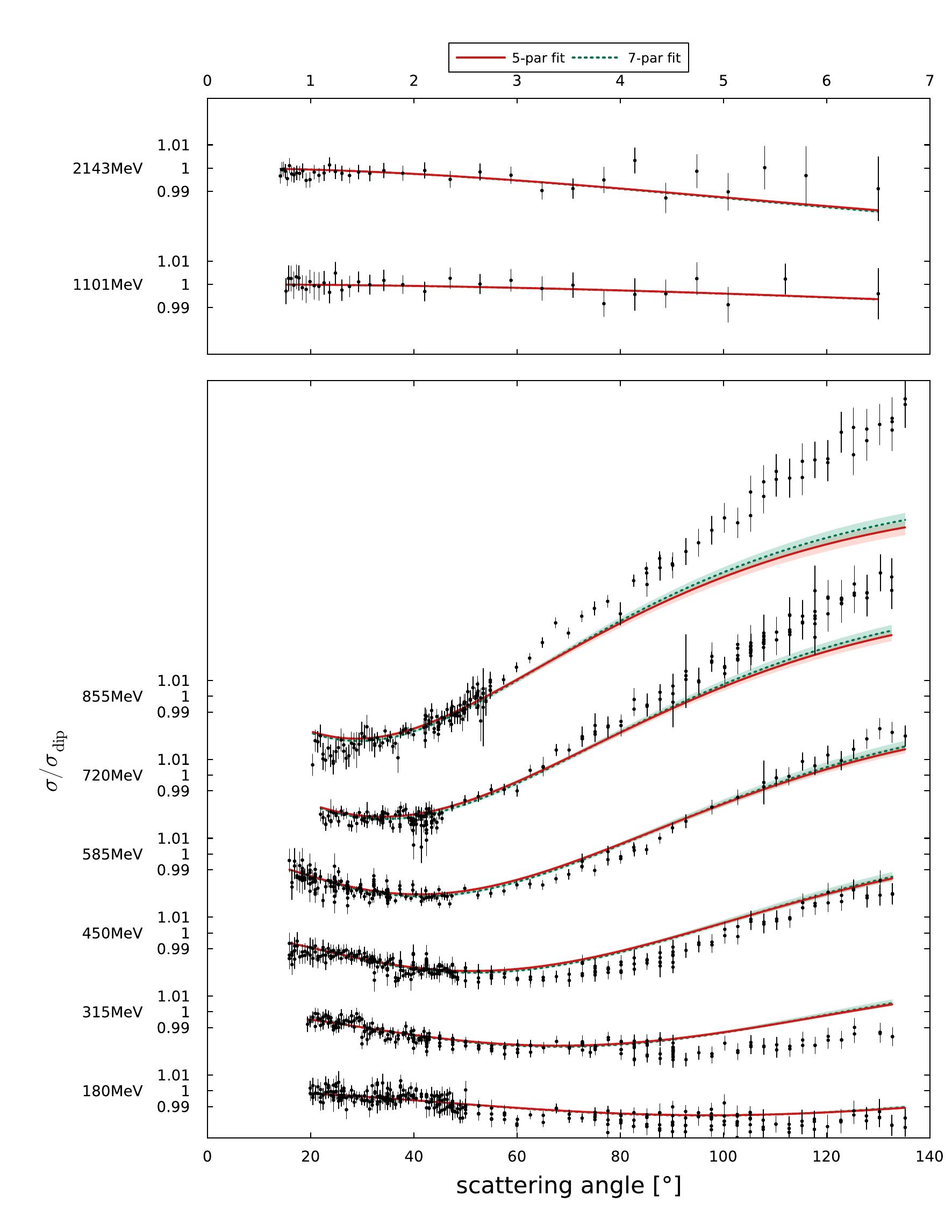} 
\caption{Fitted cross sections for the best fit compared with  data from MAMI~\cite{A1:2013fsc}
and the  data from  JLab~\cite{Xiong:2019umf}. The uncertainties are displayed by the red bands. The dashed blue lines show the best fit with equal weights
  and the dashed-dotted balck lines the fit without $G_{E,M}^p(Q^2)$ and weights as in the best fit.
  Red/green color refers to scenario~1/2.}
\label{fig:cross-sections}
\end{figure}

In Fig.~\ref{fig:ffs}, we show the
nucleon form-factors (normalized to the conventional dipole fit except for $G_E^n(Q^2)$)
for our best fit. We observe an overall good description of these data, with some
deviations due to the not overall consistent data base. 
In addition,  the form-factor ratio $\mu_P G_E^p(Q^2)/G_M^p(Q^2)$ is displayed.
As also noted in Ref.~\cite{Lin:2021xrc}, the fit does not lead to a zero crossing, but given the 
large error bars this requires further experimental scrutiny.
Also shown is the
much less constrained form-factor ratio  $\mu_n G_E^n(Q^2)/G_M^n(Q^2)$. In fact, these ratios
of the proton and neutron form-factors were first considered in a variant of the GK model
in Ref.~\cite{Lomon:2002jx}, which also does not give any zero crossing for the the proton ratio.
Its is intriguing how well this rather simple approach can describe this large number of data points.
This is also reflected in the rather good description of the differential cross section data from
MAMI (with the exception of the backward angle data for the two highest energies)
and Jefferson Laboratory, as shown in Fig.~\ref{fig:cross-sections}. Note that
in these fits, we put more weight on the form-factor data so that the MAMI data are less well
described as compared to the fit in Sect.~\ref{sec:mami}.
Specifically, we take $w_{\mu_p G_E^p/G_M^p}=w_{G_E^n}=w_{G_M^n/(\mu_n G_{\text{dip}})}=2$, 
while all other weights are set
to unity. This choice ensures a reasonably good description of the neutron form-factor data as well as the
 measurements of the proton form-factor and the neutron form-factor ratios.

The various electromagnetic radii can also be extracted from the slopes of the Sachs form-factors.
We obtain: 
\begin{eqnarray}\label{eq:radii}
{\rm Scenario~1} &:&\nonumber\\
r_E^p &=& 0.844(1)~{\rm fm}~, ~~~r_M^p = 0.829(1)~{\rm fm}~, \nonumber\\
(r_E^n)^2 &=& -0.145(6)~{\rm fm}^2~, r_M^n = 0.864(4)~{\rm fm}~.\nonumber\\
{\rm Scenario~2} &:&\nonumber\\
r_E^p &=& 0.844(1)~{\rm fm}~, ~~~r_M^p = 0.828(2)~{\rm fm}~, \nonumber\\
(r_E^n)^2 &=& -0.126(9)~{\rm fm}^2~, r_M^n = 0.854(7)~{\rm fm}~.\nonumber\\
&&
\end{eqnarray}
The proton charge radius  is still consistent with the so-called
``small value'' and the magnetic radius comes out a markedely smaller
than in dispersion theory~\cite{Lin:2021xrc}.  Also, the squared neutron electric radius comes out  
visibly larger than the best determination
from chiral effective
field theory, $(r_E^n)^2 =-0.105^{+0.005}_{-0.006}\,$fm$^2$~\cite{Filin:2020tcs}
in scenario~1 but is consistent with it in when the isoscalar spectral function is improved (scenario~2).
To the contrary,
the neutron magnetic
radius is consistent within the one  obtained using dispersion theory~\cite{Lin:2021xrc}. 

In contrast to the original GK model,  where $\Lambda_1$ comes out very different from
$\Lambda_2$ as discussed
above, in the present improved miminal value we have only one high-energy scale $\Lambda =1.49\,$GeV
and $\Lambda =1.43\,$GeV for the scenario~1 and scenario~2, respectively, 
well above the vector meson masses. This  is physically
appealling, because it points towards a  universal separation of the VMD physics from the direct quark couplings.
It is also interesting to note that converting the value of the $\Lambda$ parameter into a length scale,
namely $r_c = \sqrt{12}/\Lambda$ leads to $r_c \simeq 0.46\,$fm and $0.48\,$fm, which  agrees 
with  the universal nucleon core radius recently discussed in Ref.~\cite{Kaiser:2024vbc}.

We also performed fits with equal weights, all set to unity. The corresponding form-factors and differential
cross sections for the best fit are shown in the dashed lines in Figs.~\ref{fig:ffs},\ref{fig:cross-sections}.
The only visible differences appear in the neutron form-factors and the neutron form-factor ratio. The resulting
radii are very similar to the ones given in Eq.~\eqref{eq:radii}, the ones for the proton change be less than
0.2\%, the magnetic neutron one by 0.6\% and the electric neutron radius squared by 8\%. For these fits, no
uncertainty analsis was performed.

Finally, we performed fits similar to what was done in the dispersion-theoretical approach, that is leaving
out the data for $G_{E,M}^p(Q^2)$ from the fit. The corresponding form-factors and differential
cross sections for the best fit are shown in the dashed-dotted lines in Figs.~\ref{fig:ffs},\ref{fig:cross-sections}.
Note that the resulting proton form-factors are almost 
indistinguishable from the best fit, the most visible differences appear in the form-factor rations.
The radii are almost identical to the ones given in  Eq.~\eqref{eq:radii},
and the $\Lambda$-parameter is $\Lambda =1.48\,$GeV. Note also that the $\chi^2/{\rm dof} = 2.22$ is improved
compared to the best fit, which is not surprising.

\section{Discussion}
\label{sec:discussion}

The fit results lead to a clear physical interpretation. Once the two-pion continuum is included explicitly in the isovector  channels, an additional isolated rho-pole contribution is no longer required. This does not mean that the $\rho$ meson is physically absent. Rather, the $\rho$ appears as the dominant resonant structure inside the correlated two-pion spectral function. In the original GK model, this physics had to be represented by an effective $\rho$-pole term, because the explicit dispersive two-pion continuum was not available in a directly usable form at that time.

The present fits therefore support the interpretation
\begin{equation}
\rho_{\mathrm{eff}}\quad\longrightarrow\quad \twopi\ \mathrm{continuum},
\label{eq:rho-eff-to-2pi}
\end{equation}
where the physical $\rho$ dynamics is already contained in the dispersive two-pion input. Adding an additional isolated $\rho$ pole on top of the explicit continuum is not required and would lead to  double counting the same isovector spectral strength. In addition,
the explicit representation of the two-pion continuum allows one to include the isospin-violating $\rho$-$\omega$ mixing in a natural way.

Similarly in the isoscalar spectral function, we have
\begin{equation}
\phi_{\mathrm{eff}}\quad\longrightarrow\quad K\bar{K}\ \mathrm{continuum},
\label{eq:rho-eff-to-2pi}
\end{equation}
where the physical $\phi$ dynamics is contained in the $K\bar{K}$ contribution. However, in contrast to the isovector case,
we have an additional contribution from correlated $\rho\pi$ exchange that suppresses the pure $\phi$ dynamics.
This is very different from the original GK model, that only featured the $\phi$-meson.

The direct coupling terms remain essential. The two-pion continuum resolves the low-mass isovector strength, but it does not by itself provide the correct high-\Qsq{} behavior. The direct terms represent the nonresonant quark-current component. Together with the intrinsic GK form-factors, they provide the mechanism by which the form-factors approach the perturbative-QCD scaling regime.

The diagram in Fig.~\ref{fig:direct-coupling} illustrates one possible interpretation of this direct coupling. The photon-induced $\qqbar$ configuration is not kept as a coherent vector-meson pole, but is opened and coupled directly to the valence-quark structure of the nucleon. This is conceptually analogous to the treatment of the two-pion continuum: an effective pole-like description is replaced by a more explicit representation of the underlying dynamics.

Thus the present extension refines, rather than replaces, the GK picture. The original model interpolated between meson dynamics and quark-current dynamics. The new version resolves the effective isovector $\rho$ contribution into the physical two-pion continuum, while retaining the direct quark-current component required for the asymptotic behavior. Similarly, the $\phi$-pole of the original GK model
is substituted by the contributions from the $K\bar K$ and $\rho\pi$ continua.

\section{Conclusions}
\label{sec:conclusions}

The explicit inclusion of the two-pion continuum gives a more physical description of the isovector form-factors within the GK framework. The main conclusions are:
\begin{itemize}
\item The physical low-mass isovector strength is carried predominantly by the two-pion continuum. 
\item The $\rho$ meson remains physically present as the dominant resonant enhancement inside the two-pion spectral function.
\item In the isoscalar channel, the $\phi$-meson is part of the $K\bar K$ continuum. In addition, we have the
correlated $\rho\pi$ exchange, that partly cancels the $K\bar K$ contribution.
\item The direct coupling terms remain essential for obtaining the correct perturbative-QCD asymptotics.
\item In this improved and minimal GK ansatz, we find that we can describe the large amout of data with just
five parameters. In particular, for the first time with this type of model the high-quality data from 
Mainz and Jefferson Laboratory were included in the fit.
\item In particular, the fit requires only one high-energy scale $\Lambda$. Converting it to a core radius, we
obtain $r_c \simeq 0.48\,$fm.
\item For the first time within this type of model, an uncertainty analysis was performed.
\end{itemize}
The improved  model therefore preserves the central GK idea: a semiphenomenological interpolation between meson-dominated dynamics and quark-current dynamics over the full available \Qsq{} range. Its improvement is a clearer spectral interpretation of the isovector channel through the explicit inclusion of the two-pion continuum. Another improvemement
of the model is the improved  description of the isoscalar spectral function in the
region of the $\phi$ meson, featuring correlated $\bar{K}K$ as well as $\pi \rho$ exchange.

\section*{Acknowledgements}

We thank Hans-Werner Hammer for a careful reading of the manuscript.
The work of UGM was supported in part by Deutsche Forschungsgemeinschaft (DFG) under  Germany's Excellence Strategy -- EXC 3107 -- Project-ID~533766364,  by the CAS President's International Fellowship Initiative (PIFI) under Grant Nos. 2025PD0022 and by  the European Research Council (ERC) under the European Union's Horizon 2020 research
and innovation programme (EXOTIC, grant agreement No. 101018170).
The work of YHL was supported in part by the National Science Foundation of China under Grant
No. W2543006.

\end{document}